\documentstyle[aps,epsfig,float,amstex]{revtex}
\newcommand{\be}{\begin{equation}}
\newcommand{\ee}{\end{equation}}
\newcommand{\beq}{\begin{eqnarray}}
\newcommand{\eeq}{\end{eqnarray}}
 
\begin{document}
    
\def\gC{\mbox{\boldmath $C$}}
\def\gZ{\mbox{\boldmath $Z$}}
\def\gR{\mbox{\boldmath $R$}}
\def\gN{\mbox{\boldmath $N$}}
\def\ua{\uparrow}
\def\da{\downarrow}
\def\a{\alpha}
\def\b{\beta}
\def\g{\gamma}
\def\G{\Gamma}
\def\d{\delta}
\def\D{\Delta}
\def\e{\epsilon}
\def\z{\zeta}
\def\h{\eta}
\def\th{\theta}
\def\k{\kappa}
\def\l{\lambda}
\def\L{\Lambda}
\def\m{\mu}
\def\n{\nu}
\def\x{\xi}
\def\X{\Xi}
\def\p{\pi}
\def\P{\Pi}
\def\r{\rho}
\def\s{\sigma}
\def\S{\Sigma}
\def\t{\tau}
\def\f{\phi}
\def\vf{\varphi}
\def\F{\Phi}
\def\c{\chi}
\def\w{\omega}
\def\W{\Omega}
\def\Q{\Psi}
\def\q{\psi}
\def\de{\partial}
\def\inf{\infty}
\def\ra{\rightarrow}
\def\bra{\langle}
\def\ket{\rangle}

\title{ Exact Ground State of the $2D$ Hubbard Model at Half Filling for $U=0^{+}$}
\author{Michele Cini and Gianluca Stefanucci}
\address{Istituto Nazionale di Fisica della Materia, Dipartimento di Fisica,\\
Universit\`a di Roma Tor Vergata, Via della Ricerca Scientifica, 1-00133\\
Roma, Italy}
\maketitle
\begin{abstract}
 We solve analytically the  $N\times N$ square lattice Hubbard  model 
for even $N$
at half filling and weak coupling  by a new approach. The exact ground state wave function 
  provides an intriguing and appealing picture of the antiferromagnetic order.
Like at 
strong coupling, the ground state has 
total momentum $K_{tot}=(0,0)$ and transforms as an $s$ wave 
for even $N/2$  and as a $d$ wave 
otherwise.
\end{abstract}


The $2D$ Hubbard model 
is one of the simplest descriptions of itinerant interacting electrons on a 
lattice. The Lieb theorem\cite{l} states that 
at half filling the ground state for a bipartite lattice is unique
and has spin 
$\frac{||{\cal S}_{1}|-|{\cal S}_{2}||}{2}$
where $|{\cal S}_{1}|$ 
($|{\cal S}_{2}|$) is the number of sites 
in the ${\cal S}_{1}$ (${\cal S}_{2}$) sublattice;
here and  in the following, 
$|{\cal S}|$ will be the number of elements in the set ${\cal S}$.
For a $N\times N$ (with even $N$) square 
lattice, the ground state is a singlet. 
In Ref.\cite{md} it  was shown that in the strong coupling limit 
the ground state has total 
momentum $K_{tot}=(0,0)$ and has $s$-wave or $d$-wave symmetry for 
even or odd $N/2$, respectively.

In this paper we build the exact ground state 
of the Hubbard model at half filling in the weak coupling 
limit.  We find the same quantum numbers as predicted by 
Refs.\cite{l}\cite{md}, and this 
result strongly supports the conjecture that no phase 
transition takes place for finite  values of the Coulomb interaction parameter $U$. 

Let us consider the Hubbard model with hamiltonian
\begin{equation}
H=H_{0}+W=t\sum_{\s}\sum_{\langle r,r'\rangle}c^{\dag}_{r\s}c_{r'\s}+
\sum_{r}U\hat{n}_{r\ua}\hat{n}_{r\da},\;\;\;\;U>0
\label{hamil}
\end{equation}
on a square lattice of $N\times N$ sites with periodic boundary 
conditions and  even $N$. Here $\s=\ua,\da$ is the 
spin and $r,\;r'$ the spatial  degrees of freedom of the creation and 
annihilation operators $c^{\dag}$ and $c$ respectively. We  
represent sites by $r=(i_{x},i_{y})$ with 
$i_{x},i_{y}=0,\ldots,N-1$. The sum on 
$\langle r,r' \rangle$ is over the pairs of nearest neighbor sites and 
$\hat{n}_{r\s}$  is the number operator on the site 
$r$ of spin $\s$. The point symmetry is  $C_{4v}$, the Group
of a square\cite{appendix}. We 
Fourier expand the fermion operators: $c_{k\s}=\frac{1}{N}\sum_{r}e^{ikr}c_{r\s}$,
where $k=(k_{x},k_{y})=\frac{2\p}{N}(i_{x},i_{y})$; 
then,
$H_{0}=\sum_{k}\e(k)c^{\dag}_{k\s}c_{k\s}$ where $\e(k)=
2t[\cos k_{x}+\cos k_{y}]$.  The starting point is the following property of the number operator 
$\hat{n}_{r}=c^{\dag}_{r}c_{r}$ (for the moment we omit
the spin index).

{\em Theorem:   Let ${\cal S}$ be an arbitrary set 
of plane-wave eigenstates $\{|k_{i}\ket\}$ of $H_{0}$ and
$(n_{r})_{ij}=\bra k_{i}|\hat{n}_{r}|
k_{j}\ket=\frac{1}{N^{2}}e^{i(k_{i}-k_{j})r}$  the matrix of 
$\hat{n}_{r}$ in ${\cal S}$. 
This  matrix  has 
eigenvalues $\l_{1}=\frac{|{\cal S}|}{N^{2}}$  and
$\l_{2}= \ldots =\l_{|{\cal S}|}=0$
}. 

Note that  $|{\cal S}|\leq N^{2}$; if $|{\cal S}|= N^{2}$ 
the set is complete, like the  set of all orbitals, and the theorem is 
trivial (a particle on site $r$ is the $\hat{n}_{r}$ eigenstate with eigenvalue 
1). Otherwise, if $|{\cal S}|< N^{2}$, the theorem is an immediate 
consequence of the fact\cite{u} that
\begin{equation}
    det|(n_{r})_{ij}-\l\d_{ij}|=(-\l)^{|{\cal S}|-1}
    (\frac{|{\cal S}|}{N^{2}}-\l),\;\;\;\forall r.
    \label{det}
\end{equation}

Let ${\cal S}_{hf}$ denote the set (or shell)
of the $k$ wave vectors such that $\e(k)=0$.
 These $k$ vectors 
lie on the square having  vertices $(\pm\pi,0)$ and $(0,\pm\pi)$;
one  can show that the number of solutions, that is, the dimension of the 
set ${\cal S}_{hf}$, is $|{\cal S}_{hf}|=2N-2$. 
At half filling ($N^{2}$ holes) for $U=0$, ${\cal S}_{hf}$ is half occupied. 
In the 
$S_{z}=0$ sector 
there are  $\left(\begin{array}{c} 2N-2 \\ N-1 
\end{array}\right)^{2}$   degenerate unperturbed ground state 
configurations for $N^{2}$ holes, where all $|k\ket$ orbitals such that $\e(k)<0$ are occupied.
The first-order splitting of the degeneracy is obtained by diagonalizing the 
  $W$ matrix over the unperturbed basis; like in elementary 
atomic physics, the filled shells just produce a constant shift of all 
the eigenvalues and for the rest may be ignored in first-order 
calculations.  In other terms, we consider the truncated  Hilbert space
${\cal H}$ spanned by the states of $N-1$ holes of each spin in 
${\cal S}_{hf}$, and  we want the {\it exact} ground state(s) of $W$ 
in  ${\cal H}$; by construction ${\cal H}$ is in the kernel of $H_{0}$, so the ground 
state of $W$ is the ground state of $H$ as well.
We call $W=0$ state any vector in ${\cal H}$ which also belongs to the kernel of 
$W$.
Since the lowest eigenvalue of $W$ is zero, it is evident that any $W=0$ state 
is a ground state of $H$. We want to calculate the unique
ground state of the Hubbard Hamiltonian for $U=0^{+}$ at half
filling which
is a $W=0$ singlet with $2N-2$ holes in ${\cal S}_{hf}$ (filled 
shells are understood). 

To diagonalize the {\em local } operator $W$ in closed form  we need to set up
a {\em local }  basis set of one-body states. If  ${\cal S}_{hf}$ were 
the 
complete set of plane-wave states 
$|k\ket$, the new basis would be trivially obtained by a Fourier 
transformation, but this is not the case. The question is: how can we 
define for each site $r$ the  local counterparts of $k$ states 
 using only those 
that belong to a degenerate level? 
The answer is: build  
 a  set $\{|\varphi_{\a}^{(r)}\ket\}$ of orbitals such that 
the number operator $\hat{n}_{r}$ and the Dirac characters of the point symmetry 
Group $C_{4v}$ are diagonal. Using such a basis set for the 
half-filled shell the unique 
properties of the antiferromagnetic ground state become simple and 
transparent. 
The 
eigenvectors $|\varphi_{\a}^{(0)}\ket$ of $n_{r=0}$ and those 
$|\varphi^{(r)}_{\a}\ket$ of other sites $r$ are connected by translation 
and also by a unitary transformation, or change of basis set. 
Picking $r=\hat{e}_{l}$,  $l=x$ means $\hat{e}_{l}=(1,0)$ or transfer 
by one step towards the right and $l=y$ means $\hat{e}_{l}=(0,1)$ or
transfer by one step 
upwards.
The unitary transformation   reads:
\begin{equation}
|\varphi^{(\hat{e}_{l})}_{\a}\rangle=\sum_{\b=1}^{2N-2}|\varphi_{\b}^{(0)}\rangle
  \bra\varphi_{\b}^{(0)}|\varphi^{(\hat{e}_{l})}_{\a}\rangle
\equiv \sum_{\b=1}^{2N-2}|\varphi_{\b}^{(0)}\rangle T_{l_{\b\a}}.
\label{transferT}
\end{equation}
 The transfer matrix $T_{l}$  {\em knows} all the translational and point symmetry of the 
 system, and will turn out to be very special.

For large $N$, to find $\{|\varphi_{\a}^{(r)}\ket\}$ 
it is convenient to separate the $k$'s of ${\cal S}_{hf}$ in 
 irreducible 
representations {\em (irreps)} of the space Group\cite{hamer} $\mathbf{G}$$=
C_{4v}\otimes T$; here $T$ is the Abelian Group of the translations 
and $\otimes$ means the semidirect product. Choosing an arbitrary $k\in {\cal 
S}_{hf}$ with $k_{x}\geq k_{y}\geq 0$, the set of vectors $R_{i}k$,  
where $R_{i}\in C_{4v}$, is a (translationally invariant) basis for an irrep of $\mathbf{G}$; 
the {\em accidental} degeneracy of several irreps is due to the 
presence of extra symmetry, i.e.  $\mathbf{G}$ is a subgroup of the 
Optimal Group defined in \cite{noialtri},\cite{noialtri2}. The high
symmetry vectors $(0,\pi)$ and $(\pi,0)$ 
always trasform among themselves and are
the basis of the only two-dimensional irrep of $\mathbf{G}$, which exists for any $N$.
If $N/2$ is even,  one also finds the high symmetry  wavevectors 
$k=(\pm\pi/2,\pm\pi/2)$ which mix among themselves and yield
a four-dimensional irrep.  In general, the vectors $R_{i}k$ are all 
different, so  all the other irreps of $\mathbf{G}$ have dimension 8, 
the number of operations of the point Group $C_{4v}$. 

Next, we show how to build our {\it local} basis set and derive $W=0$ 
states  for each kind of 
irreps of $\mathbf{G}$. For illustration, we will consider the 
case $N=4$; then  
${\cal S}_{hf}$ contains the bases of two irreps of $\mathbf{G}$,
 of dimensions 2 and 4. The one with basis
 $k_{A}=(\p,0),\;k_{B}=(0,\p)$ 
breaks into $A_{1}\oplus B_{1}$ in $C_{4v }$ .

The eigenstates of $(n_{r=0})_{ij}=\bra 
k_{i}|\hat{n}_{r=0}|k_{j}\ket$, with $i,j=A,B$ , 
are $|\q''_{A_{1}}\ket=\frac{1}{\sqrt{2}}(|k_{A}\ket+|k_{B}\ket)$ 
with $\lambda_{1}=1/8$
and $|\q''_{B_{1}}\ket=\frac{1}{\sqrt{2}}(|k_{A}\ket-|k_{B}\ket)$
with $\lambda_{2}=0$.
Since under translation by a lattice 
step $T_{l}$ along the $l=x,y$ direction 
$|k\ket\ra e^{ik_{l}} |k\ket$, using 
 Equation (\ref{transferT}) one finds that 
$|\q''_{A_{1}}\ket\leftrightarrow (-1)^{\th''_{l}}|\q''_{B_{1}}\ket$, 
with $\th''_{x}=1,\;\th''_{y}=0$;  so $|\q''_{A_{1}}\ket$ has 
vanishing amplitude on 
a sublattice and $|\q''_{B_{1}}\ket$ on the other. The two-body state $|\q''_{A_{1}}\ket_{\s}|\q''_{B_{1}}\ket_{-\s}$
has occupation for spin $\s$ but not for spin $-\s$ on the site $r=0$; 
under a lattice  step 
translation it flips the spin and picks up a (-1) phase factor: 
$|\q''_{A_{1}}\ket_{\s}|\q''_{B_{1}}\ket_{-\s}
\leftrightarrow 
|\q''_{B_{1}}\ket_{\s}|\q''_{A_{1}}\ket_{-\s}$; therefore it has 
double occupation nowhere and is a 
$W=0$ state ($W=0$ pair \cite{cbs1}\cite{cbs2}).

 The 4-dimensional irrep with basis 
$k_{1}=(\p/2,\p/2),\;k_{2}=(-\p/2,\p/2),\;k_{3}=(\p/2,-\p/2)\; 
k_{4}=(-\p/2,-\p/2)$    breaks into $A_{1}\oplus B_{2}\oplus 
E$ in $C_{4v }$; letting $I=1,2,3,4$ for the 
irreps $A_{1},\;B_{2},\;E_{x},\;E_{y}$  
respectively, we can  write down all the eigenvectors 
of $(n_{r=0})_{ij}=\bra k_{i}|\hat{n}_{r=0}|k_{j}\ket$,
with $i,j=1,\ldots,4$, as 
$|\q'_{I}\ket=\sum_{i=1}^{4}O'_{Ii}|k_{i}\ket$, where $O'$ is the 
following 4$\times$4 orthogonal matrix
\begin{equation}
O'=\frac{1}{2}\left[\begin{array}{rrrr}
1 & 1 & 1 & 1 \\
1 & -1 & -1 & 1 \\
1 & -1 & 1 & -1 \\
-1 & -1 & 1 & 1 \end{array}\right].
\end{equation}
The state with non-vanishing eigenvalue is again the $A_{1}$ 
eigenstate. After a little bit of algebra we have shown\cite{u} that
under 
$T_{l}$ the subspace of  $A_{1}$ and $B_{2}$ symmetry is 
exchanged with the one  of $E_{x}$ and $E_{y}$ symmetry. Thus 
we can build a 4-body eigenstate of $W$ with vanishing eigenvalue: 
$|\q'_{A_{1}}\q'_{B_{2}}\ket_{\s}|\q'_{E_{x}}\q'_{E_{y}}\ket_{-\s}$. 
As before under a lattice step translation 
this state does not change its spatial 
distribution but $\s\ra -\s$ without any phase factor:
$|\q'_{A_{1}}\q'_{B_{2}}\ket_{\s}|\q'_{E_{x}}\q'_{E_{y}}\ket_{-\s}
\leftrightarrow
|\q'_{E_{x}}\q'_{E_{y}}\ket_{\s}|\q'_{A_{1}}\q'_{B_{2}}\ket_{-\s}$.

Now we use these results to diagonalize $n_{r=0}$ 
on the whole set ${\cal S}_{hf}$ (we could have done that directly by 
diagonalizing $6 \times 6$ matrices but we wanted to show the general 
method). 
The eigenstate  of $n_{r=0}$ with nonvanishing eigenvalue always
belongs to  $A_{1}$.
The matrix $n_{r}$
has  eigenvalues $3/8$ and (5 times) $0$, as predicted by eq.(\ref{det}). 
For $r=0$ the eigenvector of occupation $3/8$ is
$|\phi^{(0)}_{1}\rangle \equiv |\phi^{(0)}_{1,A_{1}}\rangle=
\frac{1}{\sqrt{3}}|\q''_{A_{1}}\ket+\sqrt{\frac{2}{3}}|\q'_{A_{1}}\ket$. 
The other $A_{1}$ 
eigenstate of $n_{r=0}$ has 0 eigenvalue and reads:
$|\phi^{(0)}_{2}\rangle \equiv |\phi^{(0)}_{2,A_{1}}\rangle=\sqrt{\frac{2}{3}}|\q''_{A_{1}}\ket-
\frac{1}{\sqrt{3}}|\q'_{A_{1}}\ket$. 

The other eigenvectors  , whose symmetry differs from $A_{1}$, are
$|\f^{(0)}_{3}\ket \equiv |\f^{(0)}_{B_{2}}\ket=|\q'_{B_{2}}\ket$,
$|\f^{(0)}_{4}\ket \equiv |\f^{(0)}_{B_{1}}\ket=|\q''_{B_{1}}\ket$, 
$|\f^{(0)}_{5}\ket \equiv |\f^{(0)}_{E_{x}}\ket=|\q'_{E_{x}}\ket$ and 
$|\f^{(0)}_{6}\ket \equiv |\f^{(0)}_{E_{y}}\ket=|\q'_{E_{y}}\ket$.
One finds\cite{u} that the transfer matrices $T_{l}$ of 
Equation (\ref{transferT}) such that 
$|\phi^{(\hat{e}_{l})}_{I}\rangle
\equiv \sum_{J}|\phi^{(0)}_{J}\rangle T_{l_{JI}}$,
are:
\begin{equation}
T_{x}=\left[\begin{array}{rrrrrr}
0 & 0 & 0 & -\frac{1}{\sqrt{3}} & i\sqrt{\frac{2}{3}} & 0 \\
0 & 0 & 0 & -\sqrt{\frac{2}{3}} & -\frac{i}{\sqrt{3}} & 0 \\
0 & 0 & 0 & 0 & 0 & -i \\
-\frac{1}{\sqrt{3}} & -\sqrt{\frac{2}{3}} & 0 & 0 & 0 & 0 \\
i\sqrt{\frac{2}{3}} & -\frac{i}{\sqrt{3}} & 0 & 0 & 0 & 0 \\
0 & 0 & -i & 0 & 0 & 0
\end{array}\right]\;,\;\;\;\;\;\;
T_{y}=\left[\begin{array}{rrrrrr}
0 & 0 & 0 & \frac{1}{\sqrt{3}} & 0 & -i\sqrt{\frac{2}{3}} \\
0 & 0 & 0 & \sqrt{\frac{2}{3}} & 0 &  \frac{i}{\sqrt{3}}\\
0 & 0 & 0 & 0 & i & 0 \\
\frac{1}{\sqrt{3}} & \sqrt{\frac{2}{3}} & 0 & 0 & 0 & 0 \\ 
0 & 0 & i & 0 & 0 & 0 \\
-i\sqrt{\frac{2}{3}} &  \frac{i}{\sqrt{3}} & 0 & 0 & 0 & 0
\end{array}\right].
\label{transfer}
\end{equation}
The reason why this choice of the basis set is clever is now 
apparent. The local basis at any site $r$ splits into the subsets
${\cal S}_{a}=\{|\phi^{(r)}_{1,A_{1}}\rangle,|\phi^{(r)}_{2,A_{1}}\rangle,
|\phi^{(r)}_{B_{2}}\rangle\}$, and
${\cal S}_{b}=\{|\phi^{(r)}_{B_{1}}\rangle,|\phi^{(r)}_{E_{x}}\rangle,
|\phi^{(r)}_{E_{y}}\rangle\}$;
a shift by a 
lattice step sends members of ${\cal S}_{a}$ into linear combinations 
of the members of  ${\cal S}_{b}$, and conversely.

Consider the 6-body eigenstate of $H_{0}$
\begin{equation}    
   |\F_{AF}\ket_{\s}=
   |\phi^{(0)}_{1,A_{1}}\phi^{(0)}_{2,A_{1}}\phi^{(0)}_{B_{2}}\rangle_{\s}
   |\phi^{(0)}_{B_{1}}\phi^{(0)}_{E_{x}}\phi^{(0)}_{E_{y}}\rangle_{-\s} .
\end{equation}
In this state there is partial occupation of 
site $r=0$ with spin $\s$, but no double occupation. It turns out that a 
shift by a lattice step produces the transformation
\begin{equation}
  |\F_{AF}\ket_{\s}   \longleftrightarrow - |\F_{AF}\ket_{-\s}
\label{giochino}
\end{equation}
that is, a lattice step is equivalent to a spin flip, a feature that 
we call  {\em 
antiferromagnetic property}. Since the spin-flipped state is 
also free of double occupation, $|\F_{AF}\ket_{\s}$ is a $W=0$ 
eigenstate. A ground state 
which is a single determinant is a quite unusual property for an 
interacting model like this. 

Note that $|\phi^{(0)}_{1,A_{1}}\phi^{(0)}_{2,A_{1}}\ket$ is 
equivalent to $|\q''_{A_{1}}\q'_{A_{1}}\ket$, because this is just a 
unitary transformation  of the $A_{1}$ wave functions; so
 $|\F_{AF}\ket_{\s}$ can also be 
written in terms of the old  local orbitals (without any 
mix of the  local states of different irreps of ${\mathbf G}$): 
\begin{equation}
|\F_{AF}\ket_{\s}=|\q''_{A_{1}}\q'_{A_{1}}\q'_{B_{2}}\rangle_{\s}
|\q''_{B_{1}}\q'_{E_{x}}\q'_{E_{y}}\rangle_{-\s}.
\label{perirrep}
\end{equation}
 This form of the ground state lends itself to be generalised (see below).
For $N>4$, $k$ vectors arise that do not possess any special symmetry, 
the vectors $R_{i}k$ are all 
different for all $R_{i} \in C_{4v}$, and we get  eight-dimensional 
irreps of ${\mathbf G}$.  Recalling that $|{\cal S}_{hf}|=2 N-2$,
one finds that
${\cal S}_{hf}$ contains 
$N_{e}=\frac{1}{2}(\frac{N}{2}-2)$ irreps of dimension 8, one of 
dimension 4 and one of dimension 2 if $N/2$ is even and $N_{o}
=\frac{1}{2}(\frac{N}{2}-1)$ irreps of dimension 8 and one of 
dimension 2 if $N/2$ is odd.

To extend the theory to general $N$, we note that these $k$ vectors, 
since   $R_{i}k$ are all 
different, are  a basis of  the regular representation of $C_{4v}$. 
Thus, by the Burnside theorem, each of them breaks 
into $A_{1}\oplus A_{2}\oplus B_{1}\oplus B_{2}\oplus E\oplus E$;  
diagonalizing $n_{r=0}$ and the point Group characters on the 
basis of the $m$-th eight-dimensional irrep of ${\mathbf G}$ one gets 
one-body states    $|\q_{I}^{[m]}\ket$, where $I$ stands for the 
 $C_{4v}$ irrep label, $I$=$A_{1}$, $A_{2}$, $B_{1}$, $B_{2}$, $E_{x}$, $E_{y}$, 
$E^{\prime}_{x}$, $E^{\prime}_{y}$; here  we denote 
 by $E^{\prime}$ the second occourrence of the irrep $E$.
The ground state wave function in ${\cal H}$ for the half filled case is a 
generalized version of  Equation (\ref{perirrep}). For even $N/2$, we have 
\begin{equation}
|\F_{AF}\ket_{\s}\equiv|(\prod_{m=1}^{N_{e}}
\q^{[m]}_{A_{1}}\q^{[m]}_{B_{2}}\q^{[m]}_{E_{x}}\q^{[m]}_{E_{y}})
\q'_{A_{1}}\q'_{B_{2}}\q''_{A_{1}}
\ket_{\s}
|(\prod_{m=1}^{N_{e}}\q^{[m]}_{A_{2}}\q^{[m]}_{B_{1}}\q^{[m]}_{E'_{x}}
\q^{[m]}_{E'_{y}})\q'_{E_{x}}\q'_{E_{y}}\q''_{B_{1}}\ket_{-\s},
\label{detaf}
\end{equation}
with $\s=\ua,\da$. 
For odd $N/2$, 
$|\F_{AF}\ket_{\s}$ is  similar  but the maximum  $m$ is   $N_{o}$
and the $|\q'\ket$ states do not occour. $|\F_{AF}\ket_{\s}$ is a $W=0$ 
state, transforms into 
$-|\F_{AF}\ket_{-\s}$ for 
each lattice step translation and manifestly shows an 
antiferromagnetic order ({\em antiferromagnetic property}). 
Since $W$ is 
a positive semidefinite operator $|\F_{AF}\ket_{\s}$ is actually a ground state. 
In the basis 
of local states these are the only two determinantal states 
($\s=\ua,\da$) with the above 
properties. 

A few further remarks about $|\F_{AF}\ket_{\s}$ are in order.  
1) Introducing the projection operator 
$P_{S}$  on the spin $S$ subspace, one finds that  
$P_{S}|\F_{AF}\ket_{\s}\equiv|\F^{S}_{AF}\ket_{\s}\neq 0\; , \forall S=0,\ldots,N-1$. 
Then, 
$_{\s}\bra \F_{AF}|W|\F_{AF}\ket_{\s}=\sum_{S=1}^{N-1}\, 
_{\s}\bra\F^{S}_{AF}|W|\F^{S}_{AF}
\ket_{\s}=0$, and this implies that there is at least one ground state 
of $W$ in  ${\cal H}$ for each $S$.
The actual ground state of $H$ at weak coupling 
is the singlet $|\F^{0}_{AF}\ket_{\s}$.  2) 
The {\em existence} of this singlet $W=0$ ground state 
 is a direct consequence of the Lieb theorem\cite{l}. Indeed 
the maximum spin state $|\F^{N-1}_{AF}\ket_{\s}$ is trivially in the kernel
of $W$; since the 
ground state must be a singlet it should be an eigenvector of $W$ 
with vanishing eigenvalue.  3) The above results and 
Lieb's theorem imply that 
higher order effects split the ground state multiplet of $H$
and the singlet is lowest.  4) The 
  Lieb  theorem makes no assumptions concerning the lattice 
structure; adding the ingredient of the $\mathbf{G}$ symmetry we are able 
to explicitly display the wave function at weak coupling.

Using the explicit form of $P_{S=0}$ one finds that 
$P_{S=0}|\F_{AF}\ket_{\s}=-P_{S=0}|\F_{AF}\ket_{-\s}$. This 
identity allows us to study how the singlet component transforms 
under translations, reflections and rotations. We have 
found\cite{u} that the total momentum is $K_{tot}=(0,0)$. To make contact with 
Ref.\cite{md} we have also determined how it transforms under the $C_{4v}$ 
operations with respect to the center of an arbitrary placquette. 
It turns out that it is even under reflections and transforms as an 
$s$ wave if $N/2$ is even and as a $d$ wave if $N/2$ is odd. The present
approach lends itself to obtain exact results  for other 
fillings as well, but this will be shown elsewhere\cite{cps}.


\begin{references}
    

\bibitem{l} E. H. Lieb, Phys. Rev. Lett. {\bf 62} 1201 (1989).
\bibitem{md} A. Moreo and E. Dagotto, Phys. Rev. B {\bf 41}, 9488 (1990).
\bibitem{appendix} $C_{4v}$ is the symmetry Group of a square. 
It is a finite Group of 
order 8 and it contains 4 one dimensional irreps, 
$A_{1},\;A_{2},\;B_{1},\;B_{2}$, and 1 two-dimensional one called $E$. 
The table of characters is
\begin{center}
    \vspace*{0.5 cm}
\begin{tabular}{|c|c|c|c|c|c|}
\hline 
$C_{4v}$ & $\mathbf{1}$ & $C_{2}$ & $C^{(+)}_{4},\,C^{(-)}_{4}$ & 
$\s_{x},\,\s_{y}$ & 
$\s'_{x},\,\s'_{y}$ \\
\hline 
$A_{1}$ & 1 & 1 & 1 & 1 & 1 \\
\hline 
$A_{2}$ & 1 & 1 & 1 & -1 & -1 \\
\hline 
$B_{1}$ & 1 & 1 & -1 & 1 & -1 \\
\hline 
$B_{2}$ & 1 & 1 & -1 & -1 & 1 \\
\hline 
$E$ & 2 & -2 & 0 & 0 & 0 \\
\hline 
\end{tabular}
\vspace*{0.5 cm}
\end{center}
\bibitem{u} M. Cini and G. Stefanucci, to be published.
\bibitem{hamer} Morton Hamermesh, {\it Group Theory and its 
application to Physical Problems}, Addison-Wesley (Reading, 1962).
\bibitem{noialtri} Michele Cini, Adalberto Balzarotti. Raffaella 
Brunetti, Maria Gimelli and Gianluca Stefanucci, International J. of 
Modern Physics, in press.
\bibitem{noialtri2} Michele Cini, Adalberto Balzarotti. Raffaella 
Brunetti, Maria Gimelli and Gianluca Stefanucci, cond-mat/0005223.
\bibitem{cbs1} M. Cini, G. Stefanucci and A. Balzarotti,
Solid State Communication {\bf 109}, 229 (1999).
\bibitem{cbs2} M. Cini, G. Stefanucci and A. Balzarotti, 
European Physical Journal B {\bf 10}, 293 (1999).
\bibitem{cps} M. Cini, E. Perfetto and G. Stefanucci, to be published.
\end{references}
\end{document}